# Structural and electrical properties of tantalum nitride thin films fabricated by using reactive radio frequency magnetron sputtering

H. B. Nie, S. Y. Xu, S. J. Wang, L. P. You, Z. Yang[a)], C. K. Ong and J. Li

Center for Superconducting and Magnetic Materials, Institute of Engineering Science and Department of Physics, National University of Singapore, Lower Kent Ridge Road, Singapore 119260
   a) Permanent address: Research Institute of Magnetic Materials, Lanzhou University, Lanzhou 730000, P. R. China.

T. Y. F. Liew

Data Storage Institute, DSI Building, 5, Engineering Drive 1, Singapore 117608

**Abstract**

Ta-N thin film is an attractive interlayer as well as a diffusion barrier layer in $[Fe-N/Ta-N]_n$ multilayers for the application as potential write head materials in high density magnetic recording. We synthesized two series of Ta-N films on glass and Si substrates by using reactive radio frequency sputtering under 5 mtorr $Ar/N_2$ processing pressure with varied $N_2$ partial pressure, and carried out systematical characterization analyses of the films. We observed clear changes of phases in the films from metallic *bcc* Ta to a mixture of *bcc* Ta(N) and hexagonal $Ta_2N$, then sequentially to *fcc* TaN and a mixture of TaN with N-rich phases when $N_2$ partial pressure increased from 0.0% to 30%. The changes were associated with changes in the grain shapes as well as in the preferred crystalline orientation of the films from *bcc* Ta [100] to [110], then to random and finally to *fcc* TaN [111], correspondingly. It was also associated with a change in film resistivity from metallic to semiconductor-like behavior in the range of 77K-295K. The films showed typical polycrystalline textured structure with small, crystallized domains and irregular grain shapes. Clear preferred (111) stacks parallel to the substrate surface with embedded amorphous regions were observed in the film. TaN film with [111] preferred orientation and a resistivity of 6.0 mΩ·cm was obtained at 25% $N_2$ partial pressure, which may be suitable for the interlayer in $[Fe-N/Ta-N]_n$ multilayers.

## 1.    Introduction

Recently exciting breakthroughs in processing magnetic thin film media for ultrahigh recording density on rigid disk over 25-35 $Gbit/in^2$ have been made by Seagate and IBM [1-2]. This area density is by far beyond the "theoretical limit" i.e. 10 $Gbit/in^2$ which was widely accepted a few years ago. With this achievement, to obtain a recording density of 40-100 $Gbit/in^2$ using magnetic media is not just a dream in the near future. Significant progresses have also been made in the new generation of read head by using various types of GMR (giant magnetoresistive) spin valve multilayers [3-6] that fulfill the challenging requirements of reading process at high speed/frequency on high recording density media.

However, the development of write head has not matched with the fast development of media and read head yet, and thus more intensive studies are expected in this area.

In order to obtain satisfactory overwrite performance of the recording process, the write head materials should inhabit superior properties such as very high saturation flux density $B_s$ of around 20 kG, high frequency permeability $\mu'$ of order $10^3$ at $10^{2-3}$ MHz, very low coercivity $H_c$ (say, <1 Oe), thermal stability up to 400-500 ºC and very small magnetostriction. One sort of the potential candidates for the new generation write head materials are the iron base thin films such as FeN, FeTaN and FeAlN [7-12]. To reduce the thermal noise induced by eddy current and to obtain excellent high frequency performance up to $10^{2-3}$ MHz, various [Fe(Ta, Al)N/M]$_n$ multilayers have been synthesized and investigated, where M is representing for the interlayer, e.g., TaN, Ta, $SiO_2$, $Al_2O_3$ NiFe, CoZrRe, SiN, and AlN [12-22]. Among them, with TaN as interlayer, [FeTaN/TaN]$_n$ multilayers were demonstrated to have excellent magnetic properties at high frequency up to 100 MHz and high temperature stability up to 430 ºC [12].

Ta-N is a chemically inert refractory compound. Here we use Ta-N to represent various phases of Ta-N system, such as $\beta$-Ta$_2$N, $\theta$-TaN, $\eta$-TaN, $\delta$-TaN$_{1-x}$, Ta$_5$N$_6$, Ta$_4$N$_5$, Ta$_3$N$_5$, etc. [23-24]. Before Ta-N thin film is used as an interlayer in soft magnetic multilayers, it has already been applied as a stable thin film resistor of low temperature coefficient of resistivity [25-27], an excellent diffusion barrier between silicon and metal overlayers of Ni, Al and Cu [28-33], etc. For different applications, Ta-N thin films have been successfully fabricated by using several kinds of techniques such as reactive sputtering [27-29, 31, 34-39], metalorganic chemical vapor deposition (MOCVD) [40], ion beam assisted deposition (IBAD) [41-42] and electron-beam evaporation [43]. However, due to the complex phase diagram of Ta-N system, the experimental results of different research groups are not consistent satisfactorily with each other, and the growth mechanism of Ta-N film has not been fully understood yet. We believe that a better understanding of the growth mechanism of Ta-N thin films will shed light on the synthesizing of excellent TaN interlayer films in [Fe(Ta, Al)N/TaN]$_n$ multilayers as qualified write head materials for recording density of 20-40 Gbit/in$^2$ or higher. For this purpose, it is required that the TaN interlayer have relatively high resistivity, smooth surface, good diffusion barrier capability, homogeneous thickness (e.g. 3 nm), and a stable reproducibility. All these add up to a challenging task.

In this work, radio frequency (rf) reactive magnetron sputtering technique was applied to prepare the Ta-N thin films. The sputtering process is now the most main technique for processing recording media on hard disk and for GMR/spin valve materials. It has an excellent reproducibility and a large efficient deposition area. We also fabricated Fe(Ta)N thin films with the same technique. By this way, we will be able to synthesize [Fe(Ta)N/TaN]$_n$ *in situ* after we have individually and systematically investigated Fe(Ta)N and Ta-N films on their morphological, structural and electrical properties, as well as on the correlation between the deposition parameters and the quality of as-deposited films.

## 2. Experimental

The Ta-N thin films were fabricated by using reactive rf (13.56MHz) magnetron sputtering deposition technique on a Denton Vacuum Discovery-18 Deposition System with a 3-inch-diameter tantalum target (purity 99.95%). Both 1″x1″ amorphous glass and 10 mm x 10 mm (100) Si wafers were used as substrates. The substrates were cleaned prior to the deposition with in a sequence of acetone, distilled water and ethanol in an ultrasonic bath for 15, 5 and 15 minutes, respectively. For each deposition, 2-3 pieces of glass substrates and 6-8 pieces of Si substrates were mounted on a copper substrate stage at a distance of 8-14 cm from the target surface. During deposition, the substrate stage was rotated at 9 RPM to obtain homogeneous film quality and thickness. The system was first pumped to $8\times10^{-7}$ torr in about 5 hours. Then a mixture of pure Ar/$N_2$ gases (purity 99.9995%) was induced in the deposition chamber by using mass flow controllers. The total processing ambient gas pressure was kept at 5.0 mtorr for all the films, while the $N_2$ partial pressure was varied from 0.0%, to 2.5%, 5.0%, 7.5%, 10%, 15%, 20%, 25%, up to 30% for a serious Ta/Ta-N films. The rf power was kept at 200 W, corresponding to a power density of 4.3 W/cm$^2$ on the target surface. Each film was deposited for 60 minutes. Without undergoing any post-annealing process, the as-deposited films were directly measured by using various characterization techniques including X-ray diffraction (XRD), atomic force microscopy (AFM), surface profiler, high-resolution electron microscopy (HREM), four-probe method, etc.

## 3. Results and discussion

### 3.1 Thickness, deposition rate

Measured by using an Alpha-Step 500 Surface Profiler, the thickness of the films grown on glass substrates was found varying in 403-703 nm, corresponding to a deposition rate of 6.7-11.7 nm/min, almost linearly decreasing with the increase of $N_2$ partial pressure with a gradient of -0.17 nm/min per 1.0 % of $N_2$ partial pressure, as shown in Table 1 and Fig. 1. The decrease of deposition rate could be attributed to a decreasing number of opening sites on the Ta target surface for sputtering, for many sites of the target surface are occupied by nitrogen atom, $N_2$, or Ta-N compounds when $N_2$ partial pressure increases [39, 44]. However, the linear dependence of growth rate on $N_2$ partial pressure we observed is quite different from that in previous reports [34, 39]. As will be discussed in the following sections, difference is also found in changes of phases in the films. It indicates that, even under same $N_2$ partial pressure, the growth mechanism of the Ta-N films depends greatly on the deposition system and processing parameters such as rf power density, total ambient gas pressure and target-to-substrate distance.

### 3.2 Phases, crystalline structures and hardness

The crystalline structures of the as-deposited films were measured by using $\theta$-$2\theta$ XRD with a Cu $K_\alpha$ source working as 30 kV and 20 mA. Fig. 2a shows the XRD patterns of Ta/Ta-N films deposited on glass substrate at different $N_2$ partial pressures. The pure Ta film (#1) shows two very sharp (200) and (400) peaks of the *bcc* (body-centered cubic) Ta structure. When $N_2$ was induced in the deposition process at a low partial pressure, *bcc* crystalline structure was found still dominating in the Ta(N) films, where Ta(N) representing that part of nitrogen atoms have taken the interstitial position of the *bcc* Ta lattice. However, as shown in curve #2 and #3, the closely-packed (110) faces of *bcc* Ta(N) shows much higher XRD intensity than the other existing peak, (211) of *bcc* Ta(N). Other orientation of the crystalline grains may also exist, but their XRD intensities are too small to be identified. With the increase of $N_2$ partial pressure from 2.5% to 5.0%, the (110) peak becomes lower, broader, and shifts towards the low angle direction, corresponding to an expansion of the *bcc* Ta(N) lattice constant from 3.343 Å to 3.369 Å. According to some previous results, a 5% $N_2$ partial pressure in the deposition process may already result in a atomic concentration of 40-50% of nitrogen in the as-deposited Ta(N) films [34, 39]. Thus the as-indexed (110) peaks of #2 and #3 sample in Fig. 2 in fact may consist of (110) peak of the *bcc* Ta(N) and (101) peak of hexagonal Ta$_2$N, as their corresponding lattice distance, 2.338 Å and 2.323 Å, respectively, are quite close to each other.

When the N$_2$ partial pressure is increased to 7.5% and above, the XRD patterns changes again. In curves #4-#7 for N$_2$ partial pressure of 7.5% to 20%, 5 clear peaks can be identified in each curve as (111), (200), (220), (311) and (222) of the *fcc* (face-centered cubic) TaN (also referred as *δ*-TaN$_{1-x}$ [23]) phase. It means the film consists of polycrystalline TaN grains with random orientation. At N$_2$ partial pressure of 15%, the (200) peak appears much higher and sharper than the rest. In curves #7-#9, corresponding to N$_2$ partial pressures of 20%, 25% and 30%, respectively, the (111) peak becomes dominating, implying that the crystalline grains in the films are highly oriented along [111] direction. Here a transition from random orientation to [111] preferred orientation is observed. Although, at high N$_2$ partial pressure, N-rich phases e.g. Ta$_3$N$_5$, Ta$_5$N$_6$ may appear [34], here no signs of such phases in the XRD pattern (#7-#9) were observed. Nevertheless, as revealed from the resistivity measurement discussed later (see section *3.5*), N-rich phase(s) should already exist in the film prepared under 25%, 30% N$_2$ partial pressures, which attribute to the steeply increased resistivity of the film (Fig. 7).

As the position shift of the (111) peaks in curve #4-#9 was noted, we carefully performed step scans for each sample and calculated their corresponding lattice constants. Here we suppose the lattices of the crystalline grains in these films keep their *fcc* structure. The results are shown in Fig. 3a. When N$_2$ partial pressure increases from 7.5% to 30%, the corresponding *fcc* lattice constant in the film increases by 1.4 % from 4.365 Å to 4.427 Å. Compared with the lattice constant of bulk *fcc* TaN, which is 4.336 Å [23], the relative increases of the lattice constant are 0.67 % and 2.1 % at N$_2$ partial pressures of 7.5% and 30%, respectively. Since Ta-N is a defect compound system where a wide atomic percentage range of N is allowed in its stable phases [23, 24], lattice distortions can be expected in each phase. The XRD patterns in Fig. 3a clearly show a trend of lattice cell expansion associated with the increase of N$_2$ partial pressure in deposition process. However, the lattices in film #4-#9 may not keep their perfect *fcc* structure, so that the values of lattice constant calculated in Table 1 and Fig. 3b are not absolutely true.

The side view of the [111] orientated TaN lattice consists sequent stacking layers of Ta and N atom [23]. With this [111] preferred orientation, the closest-packed faces {111} of the *fcc* TaN are parallel to the film plane, and the N atoms at vacancies or interstitial positions will help to prevent atoms from diffusing through the TaN layer, thus the film will show a best performance as a barrier layer.

To support the inferences from XRD patterns, we performed measurements using auger electron spectroscopy (AES) and derived the nitrogen concentration in the Ta-N films grown under varied N$_2$ partial pressure. Figure 4 shows the nitrogen concentration in Ta-N films as a function of N$_2$ partial pressure determined by auger electron spectroscopy (AES). The atomic concentration of nitrogen in the films increased steeply when processing N$_2$ partial pressure was increased from 0.0% to 10.0%, approaching 40-50 atm%. Then nitrogen concentration began to saturate, however, kept on increasing slowly when the processing N$_2$ partial pressure was further increased. This trend has been observed previous [34, 39]. It is consistent with the changes of phases in the films observed from XRD patterns, from *bcc* Ta(N) and hexagonal Ta$_2$N, to TaN (*δ*-TaN$_{1-x}$), and finally towards N-rich phases.

On the Si (100) substrates, we observed exactly the same trends of changes in the film structure as N$_2$ partial pressure increases. The XRD patterns are shown in Fig. 2b, and the change of responding lattice constant is shown in Fig. 3b. They are very similar to those obtained from the films grown on glass substrate. The reason is that, since we fabricated the films at room temperature, both glass and Si substrates just serve as smooth and flat surface for the films and the inter diffusion between substrate and Ta-N film could be neglected. If the films are grown at high temperature, we may expect serious diffusion effect of Si into the film thus different structures of the as-deposited films on glass and Si substrates.

Two points should be mentioned here. Firstly, the XRD patterns (Fig. 2a, 2b) indicate that the films are not well crystallized. A considerable amount of amorphous Ta-N materials is expected in the film, where crystalline grains are embedded. This kind of structure is revealed in our TEM analysis. Secondly, under the given rf power, we note that the cubic TaN phase can be easily formed in the as-deposited film when suitable $N_2$ partial pressure is applied. In a bulk Ta-N system prepared by heating Ta in high pressure of $N_2$, metastable cubic TaN can be obtained only at very high temperature, e.g., 1700 °C [45]. This implies that, in sputtering process, the energy needed for forming the cubic TaN phase could be obtained from the high energetic ions in the plasma generated by the rf or dc power [43].

In addition, the hardness of Ta-N films may be a key concern for certain application, and it also offers some indirect information on the film structure. Fig. 5 shows the results obtained from the Ta-N thin films grown on silicon, tested by the continuous stiffness measurement technique on a MTS Nanoindenter XP system with a Berkovich diamond tip. The indentation experiment was done at a constant strain rate to a depth of 350 nm. A series of 4 indentations were conducted on each sample and the results were averaged for each sample. With $N_2$ partial pressure increasing from 0.0% to 30.0%, the hardness varies between 15 GPa and 30 GPa. The highest value of hardness, 27.8 GPa, was observed in the film grown under $N_2$ partial pressure of 10.0%. The results are consistent with those reported by other researchers [46, 35].

### 3.3 Surface morphology, grain size and roughness

The surface morphologies of the films investigated by using AFM are consistent with the XRD results. AFM images were obtained on a Digital Instruments Nanoscope IIIa system with a $Si_3N_4$ cantilever operating under tapping mode. Typical AFM micrographs of the films on glass substrates are shown in Fig. 6. The grain shape shows a sequence of round, shuttle-like, and round again at $N_2$ partial pressures of 0.0%, 2.5% and 5.0%, 7.5% or higher, respectively. Obviously, the changes of the grain shape are correlated with the changes of phase in the films, i.e., from [100] oriented pure Ta film, to the mixture of [110] preferred Ta(N) and $Ta_2N$ films, and finally to random oriented or [111] preferred TaN and N-rich Ta-N films, respectively, as we observed from the XRD analyses. At $N_2$ partial pressure of 7.5%, 10%, 15%, 20%, 25% and 30%, the average grain sizes are found to be around 38 nm, 32 nm, 40 nm, 51 nm, 44 nm and 36 nm, respectively. However, their root mean square roughness calculated from a 1 $\mu$m x 1 $\mu$m area of the film does not change a lot and does not show a clear trend with the increases of film thickness, as shown in Fig. 7.

### 3.4 HREM study and microstructure

Fig. 8 shows TEM cross-section images of a Ta-N sample with thickness of 430 nm grown on (100) Si wafer under $N_2$ partial pressure of 25%. The film is quite uniform in thickness and has a smooth surface (Fig. 8a). At the Ta-N/Si interface, a thin layer of amorphous $SiO_x$ with thickness of 1.5-2.0 nm can be seen (Fig. 8b). Compared to the perfect Si lattices at the left side of the interface, the Ta-N layer on the right side does not show lattice structure as they are not epitaxial on the Si wafer. However, we observe textured structure in the Ta-N layer with clear stacks, though not perfect, parallel to the Si wafer surface, as shown in Fig. 9. Amorphous zones are also found in the Ta-N layer, which are embedded in the textured film. The average spacing of the stacks is 2.5-2.6 Å, consistent well with the TaN (111) spacing revealed in the XRD data of the film. It confirms that the only one preferred orientation of the film is [111] perpendicular to the substrate surface.

To reveal the grain size and the in-plane orientation of the crystalline structure of the film, we also investigated the plane-view of the sample. Fig. 10a is a TEM dark-field plane-view image of the sample, where clear grains with dimension of several tens of nm can be seen. However, there grains have irregular shapes. In each grain, we can observe many small, crystallized domains in the size of several nm, and probably amorphous materials between the crystallized domains. Its selected-area electron diffraction (SAED) pattern shows typical rings of polycrystalline structure of the film (Fig. 10b), indicating that the crystalline domains in the film are randomly oriented. This is further confirmed in the HRTEM images of the plane-view sample, which reveal local crystallized zones with sizes of a few nm to ten nm with varying spacing and orientation, as typically shown in Fig. 11. The TEM analyses show that, there is a preferred (111) stacking textured structure in the TaN film, which is parallel to the substrate surface. The crystallized domains in the film are small (a few to ten nm), and their in-plane orientations are random. In between the crystallized domains, there are amorphous materials. Although grain boundaries can be seen clearly, the grains do not have regular shapes.

*3.5    Resistivity*

The resistivity $\rho$ of the as-deposited films (Fig. 12) was determined by using standard four-point method with a temperature controllable testing device for the range of 77K (in liquid nitrogen) to 295 K (room temperature). At room temperature, the resistivity of pure Ta film was measured to be 95 $\mu\Omega\cdot cm$. When $N_2$ partial pressure is 2.5% or 5.0%, the film still shows a metallic behavior, i.e., a decreasing resistivity with the decrease of temperature from 295K to 77K, as typically shown in Fig. 13. On the other hand, the resistivity increases when $N_2$ partial pressure is higher than 7.5% with a semiconductor-like behavior, i.e., decreasing with the decrease of temperature (Fig. 13). We therefore argue that, when prepared under 2.5% or 5.0% $N_2$ partial pressure, although there may be a considerable $Ta_2N$ phases in the film, metallic Ta and Ta(N) still contribute greatly for the conductivity of the film, which results in a resistivity close to that of pure Ta film. But when $N_2$ partial pressure is close to 7.5%, the conducting property of the film changes characteristically. We can also tell this from the change of grain shapes from shuttle-like to round shape, as shown in Fig. 5. In the previous sections, we have discussed a clear change in phases in the films from a mixture of metallic *bcc* Ta(N) and hexagonal $Ta_2N$ to *fcc* TaN when $N_2$ partial pressure was increased from 5.0% to 7.5%. Now we note that this change was also associated with a change in the temperature dependence of resistivity from metallic to semiconductor-like behavior. Between 5.0% and 7.5%, an optimum $N_2$ partial pressure should exist, where the resistivity of the film may almost keeps constant under changing temperature. Such a film is an excellent candidate for the application in film resistor.

**4.    Conclusion**

Reactive rf sputtering is an efficient technique for obtaining various Ta-N thin films with different phases. It is found that structural, morphological and electrical properties of the Ta-N films have a strong dependence on the $N_2$ partial pressure in the deposition process of rf sputtering. In the as-deposited films grown on both glass and Si (100) substrates, clear changes in the dominating phases were observed from metallic *bcc* Ta to a mixture of *bcc* Ta(N) and hexagonal $Ta_2N$ when a small amount of $N_2$ (2.5% partial pressure or less) was induced in the deposition process, and sequentially to *fcc* TaN when $N_2$ partial pressure was close to 7.5% or higher. These changes are associated with the changes of crystalline orientation (with respect to the normal of the substrate surface) in the films, from the [100] preferred orientation metallic *bcc* Ta to [110] in the mixture of *bcc* Ta(N) and hexagonal $Ta_2N$, then to random orientation of *fcc* TaN, and finally to [111] preferred orientation of a mixture of *fcc* TaN and N-rich Ta-N film, respectively. The second change was associated with a change in the temperature dependence of the film resistivity from metallic to semiconductor-like behavior in the range of 77K-295K. Hardness of the films varied between 15 GPa and 30 GPa, peaked at 27.8 GPa in the film grown at $N_2$ partial pressure of 10.0%.

The general structure of the samples deposited under high $N_2$ partial pressure was found partially textured, with small crystalline domains in a few nm to tens of nm together with amorphous regions. They showed characteristic rings of polycrystalline structure in their plane-view electron diffraction patterns, and irregular grain shapes in the TEM dark-field images. In the TaN films grown under 25% and 30% $N_2$ partial pressure, a preferred (111) stacking lattices parallel to the substrate surface were clearly observed in the cross-section HREM images.

The room temperature resistivity of the films increases with the increase of $N_2$ partial pressure, reaches a value of around 6.0 mΩ·cm at 25% $N_2$ partial pressure and steeply increases to 14.8 mΩ·cm at 30% $N_2$ partial pressure. The deposition rate is found almost linearly decrease as the $N_2$ partial pressure increases, with a gradient of -0.17 nm/min per 1.0 % of $N_2$ partial pressure. However, the surface roughness of the films grown on both glass and (100) Si substrates is not sensitive with the change of $N_2$ partial pressure and always keeps at around 3-5 nm under high deposition rate (6.7-11.2nm/min) with a film thickness up to 673 nm.

We may conclude that, for the application of $[Fe(Ta, Al)N/TaN]_n$ multilayers, a $N_2$ partial pressure around 25 % is suitable for the fabrication of Ta-N interlayer, which inhabits mainly TaN phase, a relatively high resistivity (6.0 mΩ·cm) and preferred close-packed (111) faces along the substrate plane. These characteristics will enable the film to have a low eddy-current-induced loss at high frequency and serve efficiently as barrier layer in the $[Fe(Ta, Al)N/TaN]_n$ multilayers. As a result, the multilayers are expected to have excellent high frequency performance when used as write head materials. Our further work will be focused on finding the optimum process parameters for the TaN with film a thickness of 3-5 nm and a roughness < 0.3 nm. To achieve a low deposition rate (e.g., < 1 nm/min) could be one of the approaches.

**Acknowledgement**

The authors thank DSI (Data Storage Institute), National University of Singapore and NSTB for the financial support. The authors also thank Ms Tan Lea Peng and Ms Chen Geok Loo of DSI for their help in the measurement of AES and hardness of the samples.

**Table 1** List of the thickness, deposition rate, average grain size, root mean square roughness ($R_q$), the interplanar spacing of the (111) planes $d_{(111)}$, lattice constant $a$ and resistivity $\rho$ of rf sputtered Ta-N thin films grown on glass substrate under 5 mtorr Ar/$N_2$ processing pressure with varied $N_2$ partial pressure.

| $P_N$ (%) | Film thickness (nm) | Deposition rate (nm/min) | Average grain size (nm) | Roughness $R_q$ (nm) | $d_{(111)}$ (Å) | Lattice constant $a$ (Å) | **Resistivity $\rho$ (mΩ·cm)** |
|---|---|---|---|---|---|---|---|
| 0.0 | 703 | 11.7 | 46 | 3.4 | | | 0.095 |
| 2.5 | 673 | 11.2 | 68 | 3.7 | | 3.343 | 0.116 |
| 5.0 | 653 | 10.9 | 59 | 3.2 | | 3.369 | 0.254 |
| 7.5 | 619 | 10.3 | 38 | 3.5 | 2.520 | 4.365 | 0.426 |
| 10 | 583 | 9.8 | 32 | 3.5 | 2.529 | 4.380 | 0.702 |
| 15 | 543 | 9.1 | 40 | 3.7 | 2.532 | 4.385 | 2.70 |
| 20 | 505 | 8.4 | 51 | 4.5 | 2.540 | 4.399 | 2.81 |
| 25 | 448 | 7.5 | 44 | 4.6 | 2.553 | 4.422 | 5.93 |
| 30 | 403 | 6.7 | 36 | 3.4 | 2.556 | 4.427 | 14.8 |

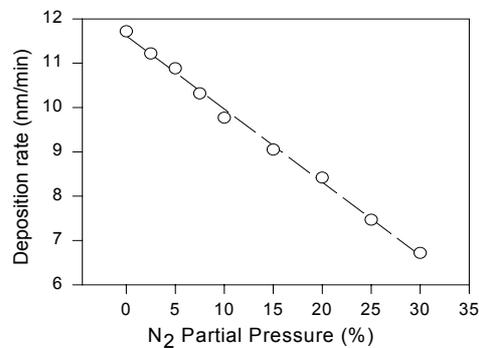

Fig. 1 Deposition rate of various Ta-N thin films grown on glass substrates vs. $N_2$ partialpressure shows a almost linear trend with a gradient of around -0.17 nm/min per 1.0 % of $N_2$ partial pressure.

**Figure 2 (a)** XRD θ-2θ patterns of the films grown on glass substrates under 5 mtorr $N_2$/Ar processing gas with various $N_2$ partial pressure of 0.0% (#1), 2.5% (#2), 5.0% (#3), 7.5% (#4), 10% (#5), 15% (#6), 20% (#7), 25% (#8) and 30% (#9).

**Figure 2 (b)** XRD θ-2θ patterns of the films grown on Si (100) substrates under 5 mtorr $N_2$/Ar processing gas with various $N_2$ partial pressure of 0.0% (#1), 2.5% (#2), 5.0% (#3), 7.5% (#4), 10% (#5), 15% (#6), 20% (#7), 25% (#8) and 30% (#9.

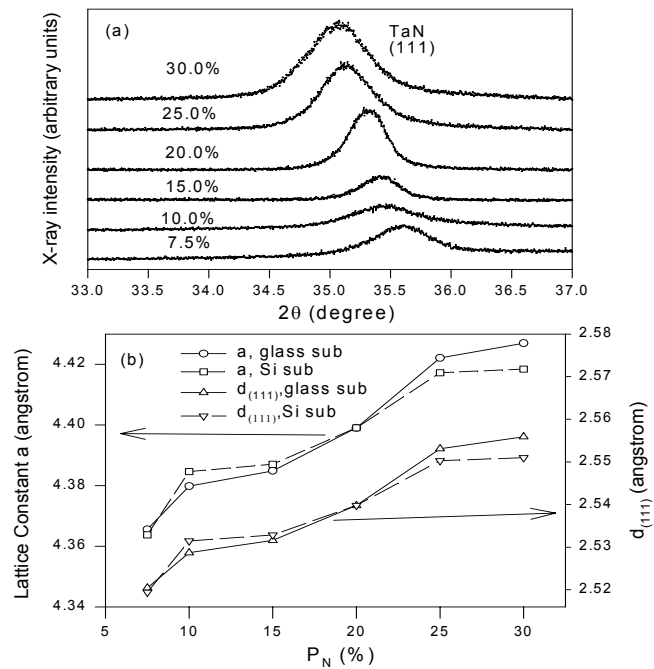

**Figure 3 (a)** The XRD step-scan patterns of the (111) peak of the curves #4-#9 in Fig. 3.2a, taken from films grown on glass substrates. The films grown on Si substrates show similar tends in their (111) step-scans. **(b)** Derived values of the interplanar spacing of the (111) planes $d_{(111)}$ and the corresponding lattice constant a of *fcc* TaN phase as a function of $P_N$ in both the films on glass and Si (100) substrates, respectively.

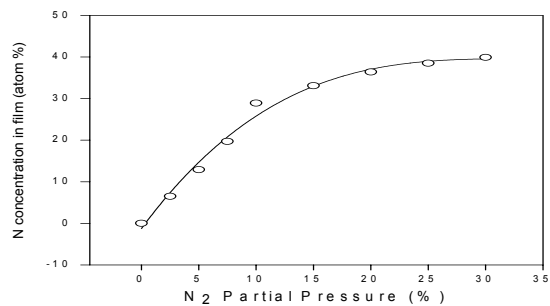

**Figure** 4 The nitrogen concentration in Ta-N films as a function of $N_2$ partial pressure determined by auger electron spectroscopy (AES).

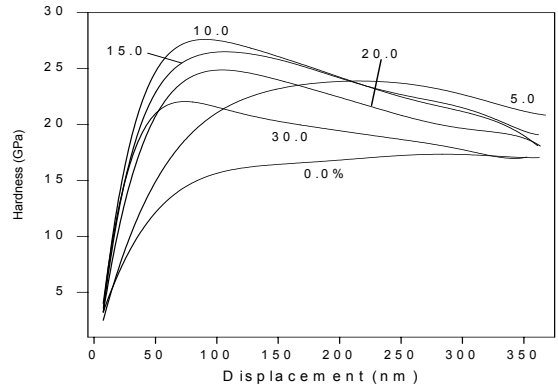

**Figure 5** The hardness of the Ta-N thin films grown on silicon substrates under varied N$_2$ partial pressure. Curves shown here are averaged from 4 individual indentations for each sample. The value of hardness peaks in the sample grown under N$_2$ partial pressure of 10.0%.

**Figure 6** The AFM tapping mode micrograghs of the Ta-N films on glass substrates, with #1-#8 corresponding to processing N$_2$ partial pressure of (a) 0.0%, (b) 2.5%, (c) 5.0%, (d) 7.5%, (e) 10%, (f) 20%, (g) 25% and (h) 30%, respectively. Each image has the same size of 0.5 μm x 0.5 μm.

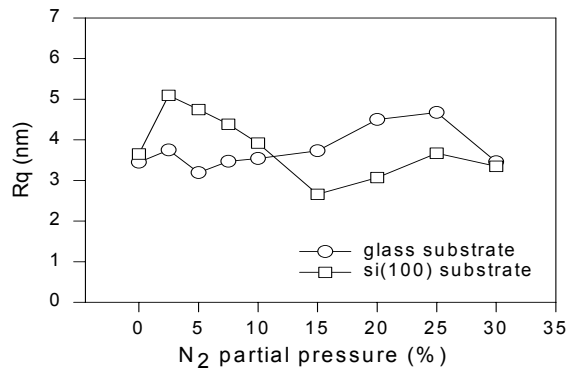

**Figure 7** The average root mean square roughness (Rq) in an area of 1 μm x 1 μm of the determined by using AFM vs. N$_2$ partial pressure.

**Figure 8** TEM cross-section images of a TaN film with thickness of 430 nm grown on (100) Si wafer under N$_2$ partial pressure of 25%. (a) Low magnification view, it shows smooth film surface and homogeneous film thickness. (b) HRTEM image of the Ta-N/Si interface shows a thin layer of amorphous SiO$_x$.

**Figure 9** HRTEM cross-section image of the Ta-N film grown on (100) Si wafer under N$_2$ partial pressure of 25% shows clear (111) stacks parallel to the substrate surface with spacing of 2.5-2.6 Å.

**Figure 10** (a) A dark-field plane-view TEM image of the Ta-N film grown on (100) Si wafer under N$_2$ partial pressure of 25% shows irregular grains with small crystalline domains in the grains. (b) A TEM selected-area electron diffraction (SAED) pattern of the sample.

**Figure 11** A typical plane-view HRTEM image of Ta-N film grown on (100) Si wafer under N$_2$ partial pressure of 25% shows various in-plane spacing of crystalline domains

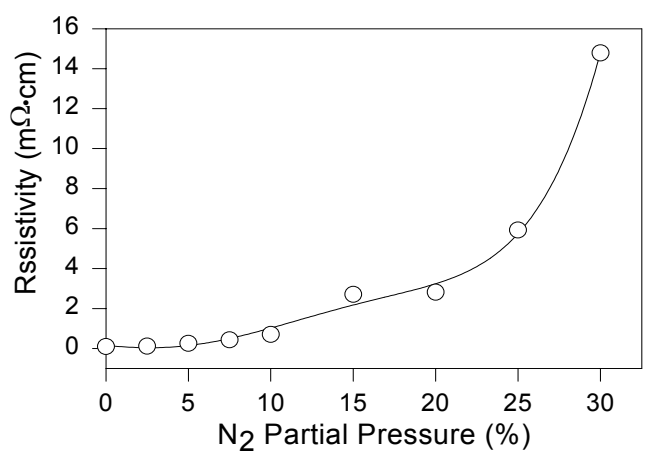

**Figure 12** The values of room temperature resistivity of Ta-N films grown on glass substrates under various N$_2$ partial pressure show a steep increase when N$_2$ partial pressure of 25%-30%.

**Figure 13** The temperature dependence of resistivity of the various Ta-N films grown on glass substrate under different N$_2$ partial pressure at 2.5%, 5.0%, 7.5%, 10%, 15.0% and 25.0 %. Note that the resistivity value has been united to $\rho(T)/\rho(295K)$, where $\rho(T)$ and $\rho(295K)$, are the resistivity at varied temperature T and 295 K, respectively.